\documentclass[a4paper]{spie}  
\usepackage{geometry}
\usepackage[]{graphicx}
\usepackage[]{graphicx}
\usepackage{array}
\usepackage{amssymb}
\usepackage{float}
\usepackage[utf8]{inputenc}  
\usepackage{amsmath,breqn}  
\usepackage{comment}
\usepackage{sidecap}
\usepackage{wrapfig}

\newcommand \mypath{{./figures}}
\newcommand{\degree}{\ensuremath{^\circ}}

\title{AGN BLR structure, luminosity and mass from combined Reverberation
Mapping and Optical Interferometry observations\footnote[2]{$\,$ Copyright 2014 Society of Photo-Optical Instrumentation Engineers. One print or electronic copy may be made for personal use only. Systematic reproduction and distribution, duplication of any material in this paper for a fee or for commercial purposes, or modification of the content of the paper are prohibited. DOI abstract:http://dx.doi.org/10.1117/12.2056436}} 

\author{Suvendu Rakshit\supit{a} and Romain G. Petrov\supit{a}
\skiplinehalf
\supit{a} Laboratoire Lagrange, UMR 7293, University of Nice Sophia-Antipolis, CNRS, Observatoire de la Cote D’Azur, BP 4229, 06304, Nice Cedex 4, France; \\
}

\authorinfo{Contact: suvendu.rakshit@oca.eu or suvenduat@gmail.com; lagrange.oca.eu}

\begin{document} 
\maketitle 

\begin{abstract}
Unveiling the structure of the Broad-Line Region (BLR) of AGN is critical to
understand the quasar phenomenon. Detail study of the geometry and kinematic of
these objects can answer the basic questions about the central BH mass,
accretion mechanism and rate, growth and evolution history. Observing the
response of the BLR clouds to continuum variations, Reverberation Mapping (RM)
provides size-luminosity and mass-luminosity relations for QSOs and Sy1 AGNs
with the goal to use these objects as standard candles and mass tags. However,
the RM size can receive different interpretations depending on the assumed
geometry and the corresponding mass depends on an unknown geometrical factor as
well on the possible confusion between local and global velocity dispersion. From RM alone, the
scatter around the mean mass is as large as a factor 3. Though BLRs are expected
to be much smaller than the current spatial resolution of large optical
interferometers (OI), we show that differential interferometry with AMBER,
GRAVITY and successors can measure the size and constrain the geometry
and kinematics on a large sample of QSOs and Sy1 AGNs. AMBER and GRAVITY (K$\sim 10.5$) could be easily extended up to K$=13$ by an external coherencer or by advanced “incoherent” data processing. Future VLTI instrument could reach K$\sim15$. This opens a large AGN BLR program intended to obtain a very accurate calibration of mass, luminosity and distance measurements from RM data which will allow using many QSOs as standard
candles and mass tags to study the general evolution of mass accretion in the Universe. This program is analyzed with our BLR model allowing predicting and
interpreting RM and OI measures together and illustrated with the results of our
observations of 3C273 with the VLTI.

\end{abstract}


\section{INTRODUCTION}
\label{sec:intro}  

The extreme power of AGN comes from the accretion of matter onto central supermassive black hole (SMBH) which is surrounded by accretion disc (AD). Broad line region (BLR), a collection of gas clouds that feeds the central source and observed in case of Sy1 AGNs, is situated somewhere between AD and clumpy dusty “Torus” which is the source of mid infrared continuum emission. The torus obscures the central part of Sy 2 objects. Detail spectroscopic monitoring, over a long time domain, of the so called “Reverberation mapping (RM)” technique allows to measure the continuum flux $C(t)$, originates from the central accretion disc region, and the line flux $L(t)$, which is the result of absorption of continuum flux and re-emission by the BLR clouds, and makes possible to estimate the size of the line emitting region by simple cross-correlation and the mass of central SMBH by a simple viral relation \cite{2004ApJ...613..682P}
\begin{equation}  
M_{bh}=f \frac{R_{blr}{\Delta V}^2}{G}, 
\label{eq: Mass}                                         
\end{equation}
where $R_{blr}=c\tau_{cent}$ is the BLR size, $\tau_{cent}$ is the centroid of the 1D response function $\Psi(\tau)$, $\Delta V$ is the width of the line, $G$ is the gravitation constant, $c$ is the speed of light and $f$ is an unknown scale factor that depends mainly on the geometry and kinematic of the object. In flat BLR geometry, the inclination of the object is the principle parameter that effect $f$ \cite{2006A&A...456...75C, 2012MNRAS.426.3086G, 2004ApJ...613..682P}.

A remarkable achievement for RM is to provide a relationship, between the size of the emission line region and the luminosity of the object, $R_{blr}\sim L^\alpha$ \cite{2013ApJ...767..149B, 2000ApJ...533..631K}. Because of the simplicity of $R_{blr}-L$ relation, it is quite easy to obtain mass-luminosity relation  $M_{bh}-L^{0.79 \pm 0.09}$ via eq.\ref{eq: Mass} \cite{2004ApJ...613..682P, 2000ApJ...528..260K, 2006ApJ...641..689V} however the scatter around this relation is quite large. The origin of this scatter is due to: A. The scatter in the size-luminosity relation which is due to inaccurate time lag and distance measurement of the individual sources \cite{2006ApJ...644..133B, 2013ApJ...767..149B}. B. Measurement of the line width $\Delta V$ also produces uncertainty in the mass estimation as the ratio of two line width measures (full width at half maximum FWHM and line dispersion $\sigma_{line}$) depend on the object and its own spectral properties.
C. The interpretation and estimation of the unknown scale factor $f$ which depends on object intrinsic properties is doubtable as using single values for time lag and line width neglect the morphology of AGN and gives large uncertainty in AGN BH mass estimation \cite{2000ApJ...533..631K,2009ApJ...697..160B}. 

For a better calibration of these relations much efforts are going on, such as:
A. Improve traditional cross correlation technique by direct modeling the driving continuum and line light curve as damped random work considering the correlated noise, de-trending and interpolation \cite{2011ApJ...735...80Z, 2009ApJ...704L..80D, 2010ApJ...721.1014M}. B. Obtain recent RM data with much higher sampling rate and free from any temporal gaps in the light curves to provide strong constraint, recover detail velocity delay map of the emission line and accurate lag measurement \cite{2012ApJ...755...60G}. C. Correct the luminosity from the contribution of host galaxy starlight for accurate distance measurement \cite{2006ApJ...644..133B, 2013ApJ...767..149B}.

After many years of effort the physics of BLR is still unknown. Detail modeling will help to improve our understanding about BLR; geometry and kinematics. Ref.~\citenum{2012ApJ...754...49P} modeled directly the RM data to recover the parameters and their uncertainties however resolving these objects spatially with high angular observing technique like interferometry in near IR or optical can constrain the geometry and morphology and calibrate the RM technique \cite{2000SPIE.4006...68P} and differential interferometry \cite{1989dli..conf..249P} could be very useful for this task \cite{2003Ap&SS.286..245M}. Ref.~\citenum{2003Ap&SS.286..261K} suggested that interferometric measurements of BLR size of quasars allows directly determination of geometrical distances on cosmic scales and long baseline ground based optical and near IR interferometer, like VLTI/AMBER \cite{2007A&A...464....1P}, has the potential of measuring the size of BELR with very high spatial resolution \cite{2003Ap&SS.286..261K}. 

 
The outline of this work is as follows. In §\ref{sec:interferometery} we described the optical interferometry and its measures with an emphasis on the observables of differential interferometry and a description of the signal and noise in OI. In §\ref{sec:Model} we described our kinematics model to explain OI and RM measures. Feasibility of present and future optical interferometric observation of the BLR of AGNs with current and future VLTI instrument and the signal to noise ratio needed for detail statistic is explained in §\ref{sec:present and future}. Our discussion is in §\ref{sec:diss} and conclusion is in §\ref{sec:conclusion}.

\section{Optical interferometry}\label{sec:interferometery}

Optical interferometry is intended to provide high angular resolution information that makes possible to study inner part of the object with very details. Modern interferometers like CHARA and VLTI with longer baselines can provide resolution down to 0.1 mas. The resolution necessary to access the dusty region of AGN is within the capability of current interferometers in IR with 8-10 m class telescopes. Since 2004 \cite{2004Natur.429...47J} more than 45 AGNs have been successfully observed in the K and N bands in LR. This has constrained the size of the innermost dust torus structure and revealed its complexity. Ref.~\citenum{2013A&A...558A.149B} rejects the existence of a simple size-luminosity relation in AGNs, because the $L^{0.5}$ scaling of bright sources fails to represent properly fainter sources. Kishimoto (2014) still tries to find an unification scheme based on the idea that in low luminosity AGNs the inner torus is more shallow than in high luminosity ones, because a latitudinal radiation pressure blows away all material far from the equatorial plane in high luminosity AGNs. Thus low luminosity AGNs have much more dust clouds in the polar direction. Both the KI and the VLTI measurements, summarized in Ref.~\citenum{2012JPhCS.372a2033K}, show that in the K band, the dust torus inner rim size is fairly close to a $R_{rim} \propto L^{0.5}$ size that can be deduced from the infrared RM measures of Ref.~\citenum{2006ApJ...639...46S}, with a size excess with regard to $\propto L^{0.5}$ that increases as L decreases but remains small in the K band. In §\ref{sec:present and future} we will use the Suganuma size as a lower limit of the inner rim size to estimate the feasibility of AGN OI observations.

The observation of BLR of AGN is more difficult because it needs a resolution higher than typically 500. The common sense in OI is that such MR observations need the use of a fringe tracker freezing the fringes from the phase jitter introduced by the atmosphere and allowing longer frame times in order to overcome the detector noise limits. Since the limiting magnitude of current fringe trackers has been lower than K$=9$, the observation of BLRs has been postponed until a FT can be operated at magnitudes K$>10$, which is supposed to be the case with the 2nd generation VLTI instrument GRAVITY expected to reach K$=10.5$ in 2016. In 2011, we have developed a new observation technique, called “blind mode observation” and a new data processing method for AMBER called “FT2D integration” \cite{2012SPIE.8445E..0WP}, inspired from the data processing used in the visible on GI2T and later on the VEGA/CHARA instrument \cite{2009A&A...508.1073M}. This gave a gain in the limiting magnitude for AMBER medium resolution observations (R=1500) from K$ \sim 7.5$ to K$\sim 10.5$ and allowed the first successful observations of the BLR of the QSO 3C273. These observations are described in Ref.~\citenum{2012SPIE.8445E..0WP}. 

A second difficulty of BLR observations is that the BLR sizes expected from RM are much smaller than the telescope diffraction limit. For example, in the case of 3C273, the maximum RM radius of 570 ld given in Ref.~\citenum{2000ApJ...533..631K} correspond to 0.36 mas angular diameter, to be compared to the 3.4 mas resolution of the VLTI in the K band. As anticipated in Ref.~\citenum{2012POBeo..91...21P} and explained in the next sections, differential interferometry allows very accurate measurement of the differential visibility and phase that can constrain strongly unresolved objects. The main results on 3C273 are: 
\begin{itemize}
\item[a)] The BLR of 3C273 is much larger (radius $1500\pm 500$ ld, or angular diameter $0.9 \pm 0.3$ mas) than expected from RM (from 250 to 570 ld) in Ref.~\citenum{2000ApJ...533..631K} and actually larger than the inner rim of the dust torus ($950 \pm 400$ ld in Ref.~\citenum{2011A&A...527A.121K}).
\item[b)] The BLR structure is very far from a flat disc dominated by global velocities. The orbits of the BLR clouds are either distributed in a very large opening angle range or affected by very strong local turbulent velocities.
\end{itemize}
The success of these observations triggered the development of the modeling tools described in Rakshit et al. (2014) and summarized in this paper. The description of the 3C273 observation and their interpretation are described in another paper (Petrov, 2014) but we will use the measurement accuracies actually achieved in section §5 where we discuss the possibility to observe AGN BLRs with existing or close in the future OI instruments.

\subsection{Spectro-interferometric measurements}
An interferometer with baseline B yields the complex visibility of the source, i.e. the normalized Fourier Transform $\tilde{O}(\mathbf{u}, \lambda)$ of the source brightness distribution $O(\mathbf{u}, \lambda)$ at the spatial frequency $\mathbf{u}=\mathbf{B}/\lambda$.
\begin{equation}
	\tilde{O}(\mathbf{u}, \lambda) =\frac{\int\int{O(\mathbf{r}, \lambda)} e^{-2\pi i \mathbf{u.r}}\,\mathrm{d^2}\mathbf{r}}{\int\int{O(\mathbf{r}, \lambda)\, \mathrm{d^2}\mathbf{r}}} = V_{*}(\lambda) e^{i\phi_{*}(\lambda)},	
\label{eq:OI}
\end{equation} 
where the modulus $V_{*}(\lambda)$ of $\tilde{O}(\mathbf{u}, \lambda)$ is given by the contrast of the fringes and called source absolute visibility whereas source phase $\phi_{*}(\lambda)$ is given by the position of the fringes at the frequency $\frac{\mathbf{B}}{\lambda}$. If we have enough number of baselines (u-v points) to sample the frequency plane (u-v plane) with a step f within a diameter $B_{max}$ it is possible to reconstruct the source brightness distribution $O(\mathbf{r}, \lambda)$ with a resolution $R=\lambda/B_{max}$ within a field $F=\lambda/f$ by inverse Fourier transform of $\tilde{V}(\mathbf{u},\lambda)$.

To summarize, a spectro-interferometric instrument produces the following measurable in each spectral channel: (1) the source spectrum $s(\lambda)$ which is deduced from the photometric measures, (2) the source absolute visibility $V(\lambda)$ with an uncertainty of at least 0.03 because of the need to calibrate it on a reference source, (3) the source differential visibility $V_{diff}(\lambda)$, (4) the source differential phase $\phi_{diff}(\lambda)$ and (d) the source closure phase $\Psi(\lambda)$.

The accuracy of the “self-calibrated” quantities, $V_{diff}(\lambda)$ , $\phi_{diff}(\lambda)$ and $\Psi(\lambda)$ are strongly dominated by the fundamental noise limits set by the photon noise, the detector noise and the thermal background photon noise, at least for MR observations over a small wavelength range.

\subsubsection{Differential interferometry of non-resolved sources}

A non-resolved source has a global angular size $\Lambda$ smaller than the interferometer resolution limit $\lambda/B$.
In eq.\ref{eq:OI}, this implies that $O(\mathbf{r},\lambda)$ is different from 0 only for values of $r<\lambda/B=1/u$, i.e. the integral in eq.\ref{eq:OI} can be limited to values $\mathbf{u}.\mathbf{r}<1$. Since Ref.~\citenum{1989dli..conf..249P} we know that the interferometric phase for such a source is given by
\begin{equation}
\phi_{*ij}(\lambda)=2\pi \mathbf{u}_{ij} \boldsymbol{\epsilon}_{ij}(\lambda),
\end{equation}
where the quantity
\begin{equation}
 \boldsymbol{\epsilon}_{ij}(\lambda)=\frac{\int\int{\mathbf{r}O(\mathbf{r}, \lambda)\, \mathrm{d^2}\mathbf{r}}} {\int\int{O(\mathbf{r}, \lambda)\, \mathrm{d^2}\mathbf{r}}}
 \label{eq:photo}
\end{equation}
is the photocenter of the source for single mode interferometry $u_{ij}=\frac{B_{ij}}{\lambda}$. From Appendix \ref{sec:app} we can say that:

\begin{itemize}
\item The closure phase decreases as $\alpha^3$ where $\alpha = \frac{\Lambda B}{\lambda}$ and stops to be usable very rapidly when the source gets unresolved. As the closure phase is necessary for a full image reconstruction, we see that images of BLRs are excluded if we don’t have baselines allowing to approach $\alpha=1$, i.e. at least 1 km for near IR observations.
\item The differential phase decreases only as $\alpha$, which makes it possible to measure it for very unresolved sources, given a sufficient SNR.
\item The visibility and the differential visibility decreases as $\alpha^2$. As we will see in §\ref{sec:target}, this allows measurements to some targets with the existing VLTI baselines.
\end{itemize}

\section[Dynamical model of BLR]{Dynamical model of BLR\footnote[3]{\MakeLowercase{\MakeUppercase{F}or detail description of the model and the result, see \MakeUppercase{R}akshit et al. (2014)} }}\label{sec:Model}
In order to produce simultaneously OI and RM measures we developed a model considering that the BLR consists of large number of line emitting clouds each defined by own position, intensity and velocity components. This model is described in detail in Rakshit et al. (2014) and here we just make a short summary. 

The radial position $r$ of the clouds are taken from a gaussian distribution with two parameters $R_{in}$ and width $\sigma_{blr}$ 
\begin{equation}
 l=\mathcal{N}(R_{in}, \sigma_{blr})\\
 \quad \text{and} \quad r=l \quad \text{for $ l \ge R_{in} $}
\end{equation}     
For velocity, we considered orbital, radial inflow, outflow and random turbulence components:
\begin{description}
\item[a)] An orbital component: 
\begin{equation}
V_{rot}  = V_{r}(\frac{R_{in}}{r})^{\beta},
\end{equation}
The parameter $\beta$ defines different orbital velocity components. For Keplerian motion $\beta =0.5$ and amplitude $V_{r}=\sqrt{\frac{GM_{bh}}{R_{in}}}$ 

\item[b)] A radial component (inflow or outflow): 
\begin{equation}
V_{rad} = V_{c}(\frac{R_{in}}{r})^{\gamma},
\end{equation}
where $\gamma$ is power law index of radial component of velocity. $\gamma=0.5$ is the freefall of clouds with amplitude $V_{c}=-\sqrt{\frac{2GM_{bh}}{R_{in}}}$ and $\gamma=-1$ is the outflow case with the outflow velocity amplitude $V_{c}$ set at the inner radius $R_{in}$ of the BLR.

\item[c)] A random macroturbulence velocity component $V_{turb}$ that increases the local velocity. Its amplitude depends on the local thickness of the BLR \cite{2006A&A...456...75C}.
\end{description}

 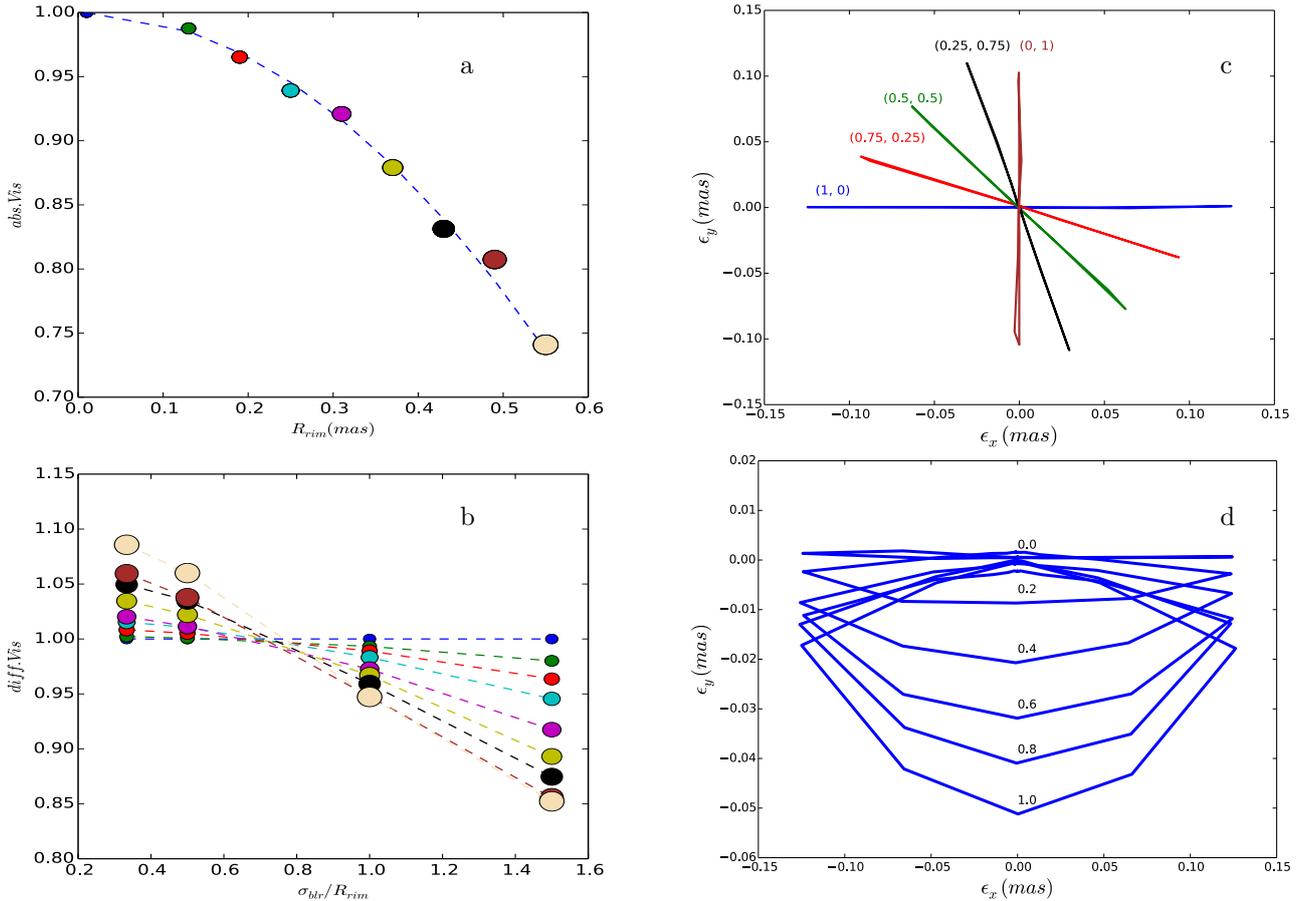
\begin{figure}[!h]
 \setlength{\unitlength}{1cm}
  \begin{picture}(18, 12)
  \put(0, 0){\includegraphics[width=8cm, height=12cm]{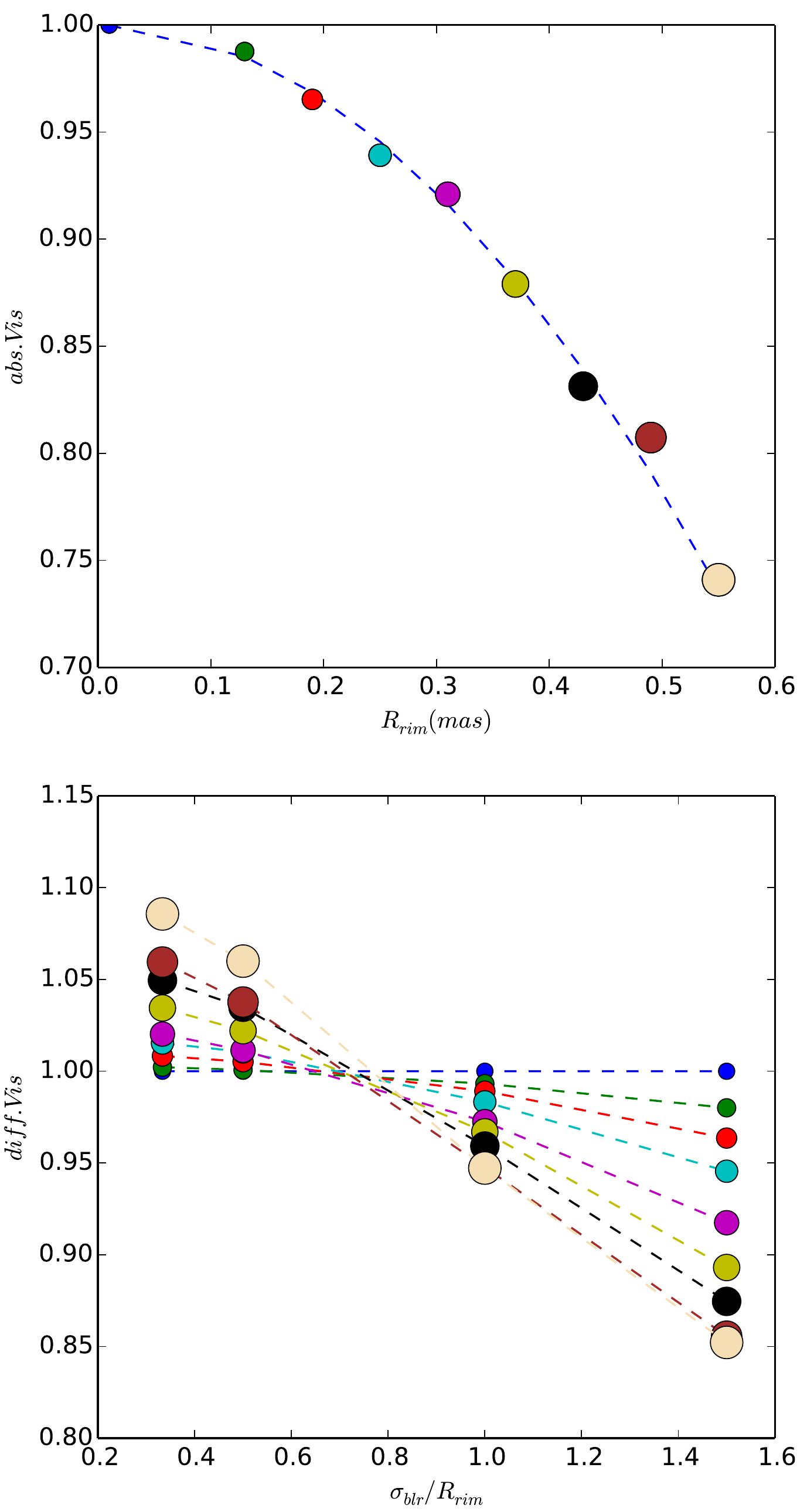}}
  \put(9, 6){\includegraphics[width=8cm,height=6cm]{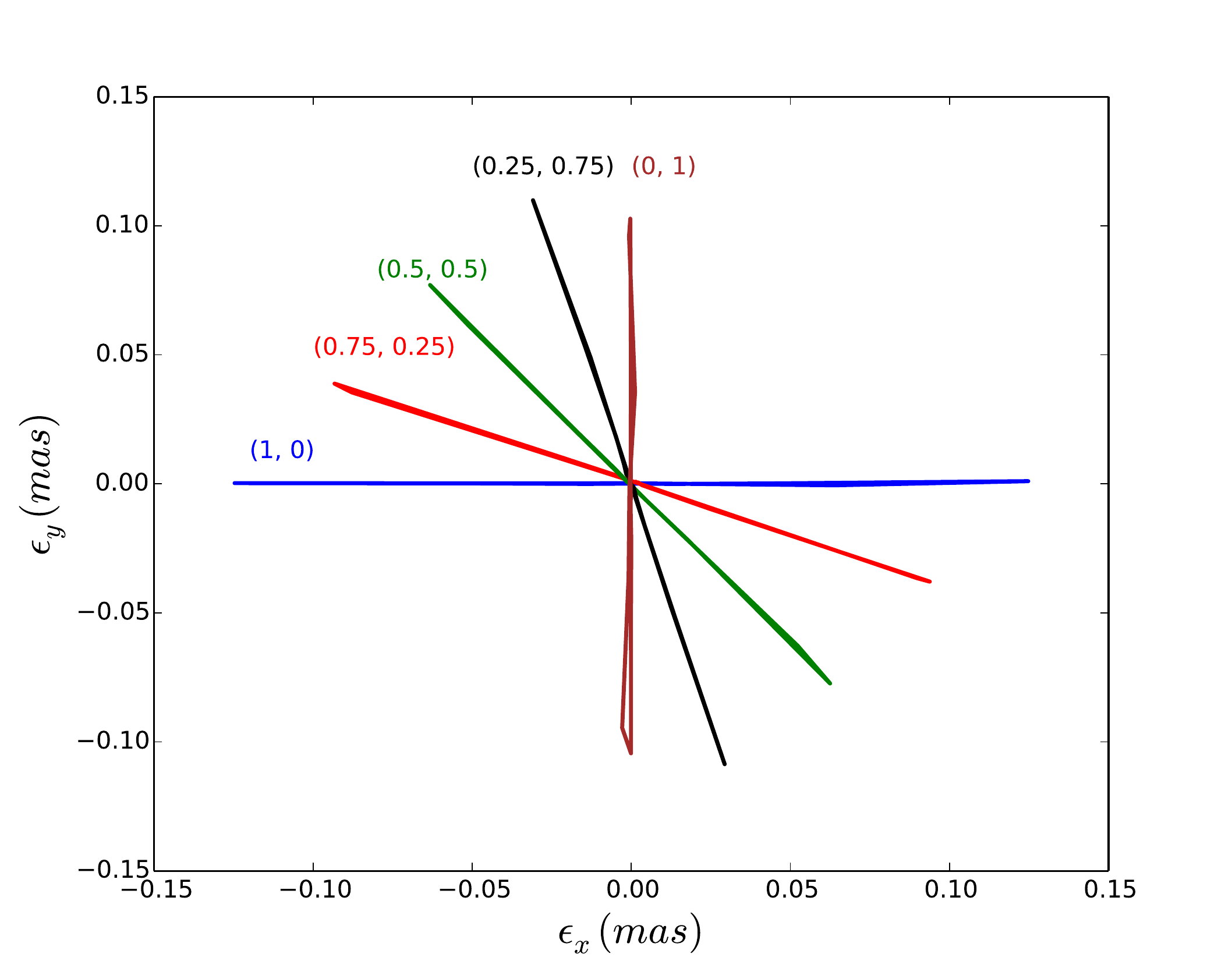}}
  \put(6.0, 5){b}
  \put(6.0, 11){a}
  \put(9,0){\includegraphics[width=8cm,height=6cm]{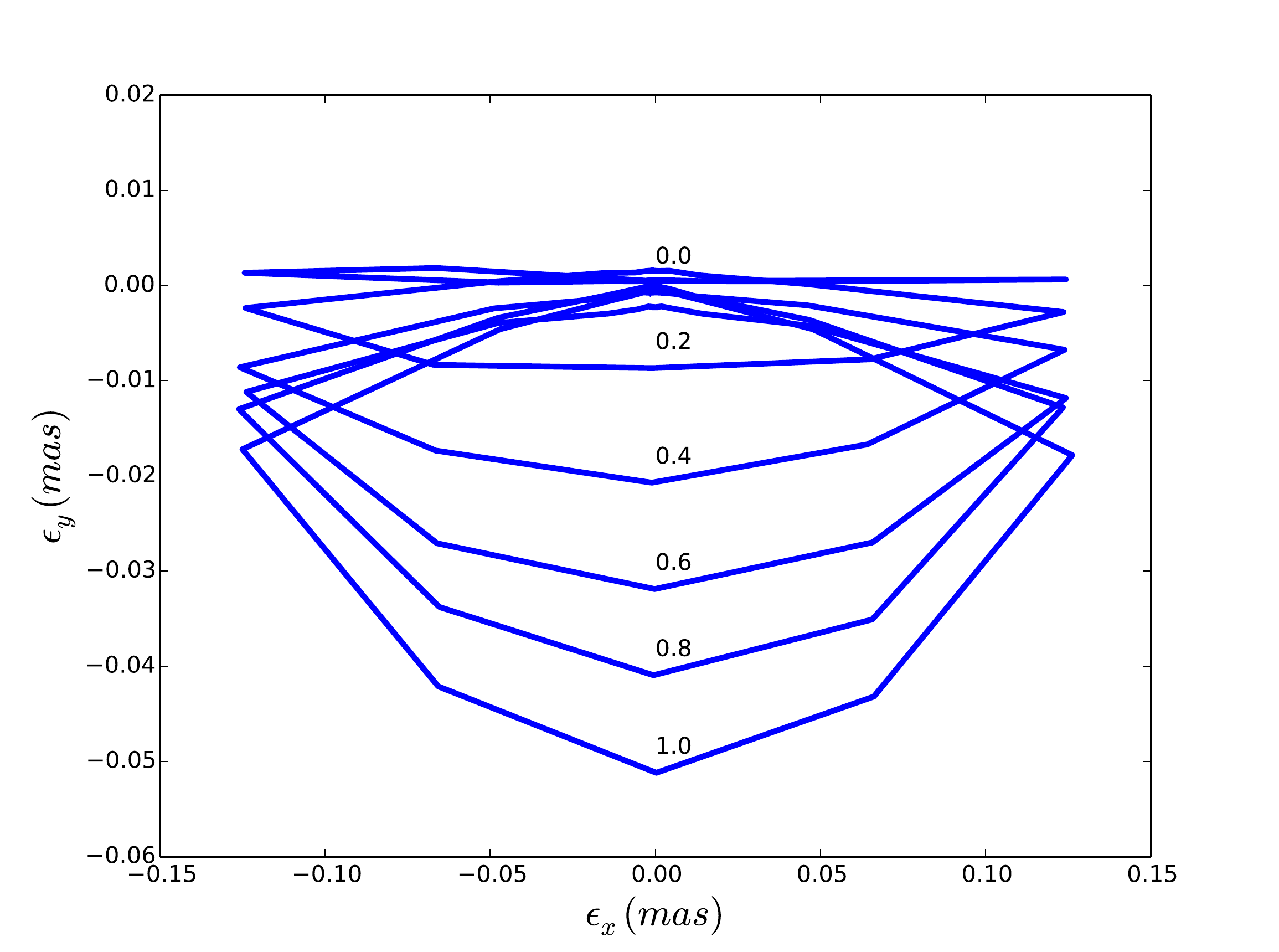}}
  \put(16.0, 5){d}
  \put(16.0, 11){c}	
  \end{picture}
 
 \caption{a. Absolute visibility in the continuum as a function of $R_{rim}$ (upper panel). b. Differential visibility as a function of $\sigma_{blr}/R_{rim}$ (lower panel) for $PA=0\degree$. c. Photocenter displacement in the sky plane for different combinations ($V_k$, $V_f$) of Keplerian ($V_k \times V_r$) and freefall velocity ($V_f \times V_c$) amplitude. The photocenter for pure Keplerian case is represented in blue whereas pure freefall is presented in brown. d. Photocenter displacement on the sky plane with anisotropy.}\label{Fig:cont}
 \end{figure} 

First, we created a 2D distribution of the clouds and then we applied: (a) A random rotation to the cloud position with angle $\omega$ ($-\omega\le\theta $(polar angle)$\le+\omega$) along y axis to pull up a 3D configuration from the 2D configuration. The angle $\omega$ is defined as opening angle or illumination angle of the cloud distribution. For $\omega = 90\degree$, the geometry become spherical whereas $\omega=0$ converge to the flat geometry. (b) A random rotation to restore the axisymmetry of the cloud distribution by removing the concentration of clouds at the intersection of the orbits i.e., $+Ymax$ and $-Ymax$.

To define the observer reference system, we rotate the clouds: (a) About y-axis with inclination angle $i$ to model the system's inclination with respect to the observer line of sight. We defined $i=0\degree$ for face-on configuration and $i=90\degree$ for edge-on configuration. (b) About z axis with an angle $PA_{disk}$ defined as the position angle of the system (from $\mathbf{N}$ to $\mathbf{E}$). The position angle $0\degree$ means the orientation of semi major axis is along North (+y axis)-South (-y axis) and usually constrained by the orientation of radio jet for QSOs.

We considered that each clouds contributes to the BLR intensity distribution by adding a spectrum which is Gaussian in shape with width $\sigma 0_{line}$ shifted by its projected line of sight velocities. Object intensity distribution is summed of the continuum and the BLR intensity distribution. We modeled continuum emission as thin ring of radius $R_{rim}$.
The main parameters that define our model are: 1. Mass of central BH: $M_{BH}$. 2. Inclination angle: $i$. 3. Opening angle: $\omega$. 4. Local line width: $\sigma0_{line}$. 5. Size of the BLR: $\sigma_{blr}$. 6. The size continuum rim: $R_{rim}$.

\subsection{BLR size estimation}
The plot in Fig.\ref{Fig:cont}a allows to discuss the accuracy on the $R_{rim}$ resulting from error in absolute visibility $\sigma_{avis}=0.03$ on the VLTI. It appears that the lower limit for possible measures is $R_{rim}=0.15$ mas. Accessing smaller $R_{rim}$ requires either longer baselines (i.e. a future interferometer) or more accurate visibilities.

Fig.\ref{Fig:cont}b illustrates the measurement of the ratio $\sigma_{blr}/R_{rim}$ from the differential visibility. We see that, even for a very small $\sigma_{dvis}$, the estimation of the size ratio is strongly dominated by the uncertainty on the continuum size, i.e. the absolute visibility accuracy. Thus differential visibility allows to say if the BLR is larger or smaller than the inner dust rim, but any accurate BLR size measurement requires an accurate absolute visibility measurement in the continuum. If $\sigma_{avis}$ is improved, then very accurate differential visibility gives access to BLR sizes much smaller than $R_{rim}$, of the order of 0.2 $R_{rim}$ for a typical value of $R_{rim}=0.2$ mas.

The second generation VLTI instrument GRAVITY is expected to allow a big improvement in the error of differential visibility, $\sigma_{dvis}$, at least in the magnitude range (up to K $\sim 10.5$) allowing to use its internal fringe tracker that will stabilize the instrument visibility.

\subsection{Kinematics}

Fig.\ref{Fig:cont}c shows the displacement of the photocenter in the sky plane for different ratios of Keplerian and freefall velocity amplitude. Global direction of photocenter displacement with respect to the rotation axis yields the rotation/expansion velocities ratio which also pointed out by Ref.~\citenum{1992ESOC...39..403C} in a similar model for circumstellar disks. In a similar way Ref.~\citenum{1996A&A...311..945S} has shown that the shape of the differential phase is the best constraint on the velocity law index $\beta$ and $\gamma$.

\subsection{Anisotropy}

The cloud line optical depth is a key factor to determine the probability of line photon escaping the cloud. The response of the emission line cloud hence depends on the anisotropic emission of the BLR. Calculation of anisotropic response for different strong broad emission line in AGN spectra \cite{1994MNRAS.268..845O} suggested that line with large ionization parameter emitted anisotropically at some radii. 
Anisotropic emission could be constrained by detail modeling of BLR clouds. To test the effect of anisotropy on the interferometric measurement we define a simple “moon phase” like anisotropy in the form of 
\begin{equation}
I(\phi)=(1-F_{anis}\cos\phi \sin i)
\end{equation}
where $F_{anis}$ is anisotropy parameter ranging between 0 (complete isotropy and optically thin cloud) to 1 (complete anisotropy and optically thick cloud).

Fig.\ref{Fig:cont}d shows the displacement of photocenter (using eq.\ref{eq:photo}) in the sky plane for flat Keplerian disk geometry. As anisotropy increases the photocenter moves towards the direction of increased emission and hence photocenter in the parallel direction changes rapidly while the photocenter in perpendicular direction remain unchanged.

  \begin{SCfigure} 
  \centering
  \resizebox{6cm}{5cm}{\includegraphics{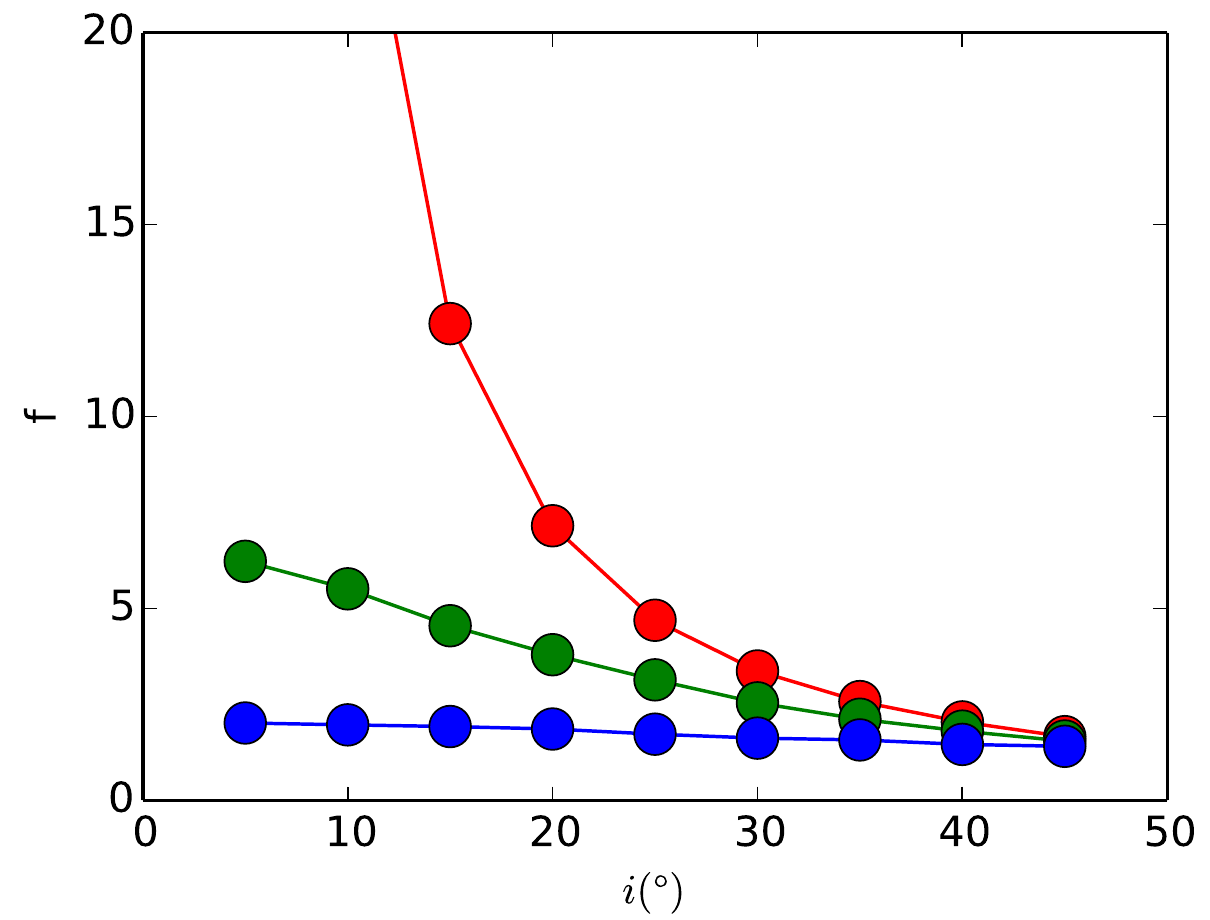}}
  \caption{The scale factor $f$ is plotted as a function of inclination for different opening angles $\omega=0\degree$ (red), $30\degree$ (green) and $60\degree$ (blue). $f$ strongly depends on $i$ and $\omega$. To constrain $f$, it is necessary to understand to geometry of the object. }\label{Fig:MassRatio}
  \end{SCfigure}

\subsection{The effect of inclination and opening on RM scale factor}\label{sec:inc}
 It is straightforward to calculate RM scale factor using the centroid of response function $\tau_{cent}$ and line dispersion $\sigma_l$ (putting $\Delta V=\sigma_l$ in eq.\ref{eq: Mass}). Fig.\ref{Fig:MassRatio} shows the virial scale factor as a function of inclination for different BLR geometries. In all cases $f>1$, which means the estimated mass is less than the mass we input in the model. As inclination increases the scale factor decreases and hence the estimated mass become closer to the input mass. For low inclination, the output mass is lower for thin geometry (small $\omega$) than the thick (larger $\omega$) and hence the scale factor is large for thin geometry than thick geometry. For higher inclination the output mass or the scale factor appear to converge with each other for different geometries. For any inclination lower than typically 30$\degree$ to 40$\degree$, $i$ and $\omega$ introduce a large $f$ uncertainty, that can be drastically reduced if we can constrain one of these parameters from other observations. The uncertainty of such estimation, from models of the jet and of the line profile is very large and though the inclination is the major factor in mass determination but currently we do not have accurate measurement of inclination of type 1 objects. Without having accurate inclination measurement it would be very difficult to constrain the scale factor.


\section{Interferometric observation of BLR}\label{sec:present and future}

In this section we will discuss the possibility to observe QSOs and Sy1 AGNs with various VLTI instruments that already exist, are under construction or could exist in a near feature. In each case we will examine:
\begin{itemize}
\item[1)] The possibility to observe the target, i.e. to detect and maintain the fringes on the target itself.
\item[2)] The accuracy of the absolute visibility, differential visibility and differential phase that can be obtained on this target.
\end{itemize}

\subsection{Interferometric signal and noise}\label{sec:ref}

From a general formalism described in Ref.~\citenum{1986JOSAA...3..634P} and updated in Ref.~\citenum{2006MNRAS.367..825V}, it is easy to show that the noise on the coherent flux computed from each multi-axial all in one interferogram is given by:
\begin{equation}
\sigma_C=\sqrt{n_Tn_*t_{DIT} + n_p\sigma_{RON}^2 + n_T n_{th}t_{DIT}},
\end{equation}
 where
  $n_*$ is the source flux per spectral channel, frame and second,
  $n_T$ is the number of telescopes,
  $t_{DIT}$ is the frame exposure time,
  $n_p$ is the number of pixels (or of measures),
  $\sigma_{RON}^2$ is the variance of the detector read-out noise and
  $n_{th}$ is the background flux per spectral channel, frame and second. In K band this value is much smaller than the detector noise and hence can be negligible for short exposures. However for long exposures such as in cophased  mode $n_{th}$ should be taken into account. In K band $n_{th}=1.07$ photons $see^{-1} cm^{-2} \mu m^{-1}$.

The classical SNR on the coherent flux, per spectral channel and per frame \cite{2006MNRAS.367..825V, 2012SPIE.8445E..2JL} is then given by (see also; MATISSE performance Analysis report, Doc No: VLT-TRE-MAT-15860-9007)
\begin{equation}
SNR_1=\frac{C}{\sigma_C} \simeq \frac{n_*t_{DIT}V}{\sigma_c},
\end{equation}
where $V$ is the visibility module.
 The source flux per spectral channel per frame and per second is given by
 \begin{equation}
 n_{\star}=n_0\mathrm{AST}\delta{\lambda}10^{-0.4\mathrm{K}},
 \end{equation}
 where $n_0$ is the number of photons per $cm^2$, $\mu m$ and sec from a star with $\mathrm{K}=0$, outside earth atmosphere, $n_0=45 \times 10^6$ photons $sec^{-1} cm^{-2} \mu m^{-1}$,
 A is the collecting area of telescope, 
 S is the Strehl ratio with the VLTI Adaptive optics system MACAO,
 T is the overall transmission of the atmosphere, the VLTI and the instrument, and
 $\delta \lambda$ is the spectral band width $=\lambda_0/\mathrm{R}$, where R is the resolution.

{\bf Standard processing:} as the coherent flux C is affected by a random atmospheric phase, we have to average its squared modulus $|C|^2$ over all available spectral channels and several frames. The SNR of such a quadratic average is given by:
 \begin{equation}
 SNR(|C|^2)=\frac{SNR_1^2}{\sqrt{1+2SNR_1^2}}\sqrt{N_{EXP}n_\lambda},
 \end{equation}
 where $n_\lambda$ is the number of spectral channels, $N_{EXP}=\frac{t_{EXP}}{t_{DIT}}$ is the total number of $t_{DIT}$ frames processed in the $t_{EXP}$ total time.
 
 {\bf AMBER+ processing:} we have developed a new approach where the full dispersed fringe image is processed, in a way equivalent to a coherent integration of all spectral channels, whatever the SNR1 per channel is. This data processing is explained in Ref.~\citenum{2012SPIE.8445E..0WP}. Then we still have to make a quadratic average of the other frames and the SNR of this processing is given by
 \begin{equation}
  SNR_{+}(|C|^2)=n_\lambda\frac{SNR_1^2}{\sqrt{1+2n_\lambda SNR_1^2}}\sqrt{N_{EXP}},
 \end{equation}
 This allowed a gain of typically $\sqrt{n_\lambda}$ which made possible the first observation of 3C273 with a spectral resolution R=1500. The fringes where detected with a typical $SNR{+}(|C|^2)=3$ in 1 s exposures.

The phase is estimated from the average coherent flux and its accuracy is given by
 \begin{equation}
 \sigma_{\phi}=\frac{<C>}{\sigma_C \sqrt{2}}=\frac{1}{SNR_1 \sqrt{2}\sqrt{N_{EXP}}}
 \end{equation}
 with $N_{EXP}=36000$ for 2 hours of observations.
 
In AMBER+, a SNR analysis (Petrov et al, 2014 SPIE) shows that 
 \begin{equation}
 \sigma_{\phi+}=\sigma_{\phi}\sqrt{2\frac{\sigma_{\phi}^2}{n_\lambda} + \frac{1+n_\lambda}{n_\lambda}}
 \label{eq:phiAMBER+}
\end{equation}
The above equations are used in §\ref{sec:present and future} to calculate the SNR on coherent and the number of target accessible with the current and upcoming VLTI instruments.

\begin{figure}[!hb]
 \setlength{\unitlength}{1cm}
 \begin{picture}(18, 7)
 \put(0, 0){\includegraphics[width=8cm, height=7.4cm]{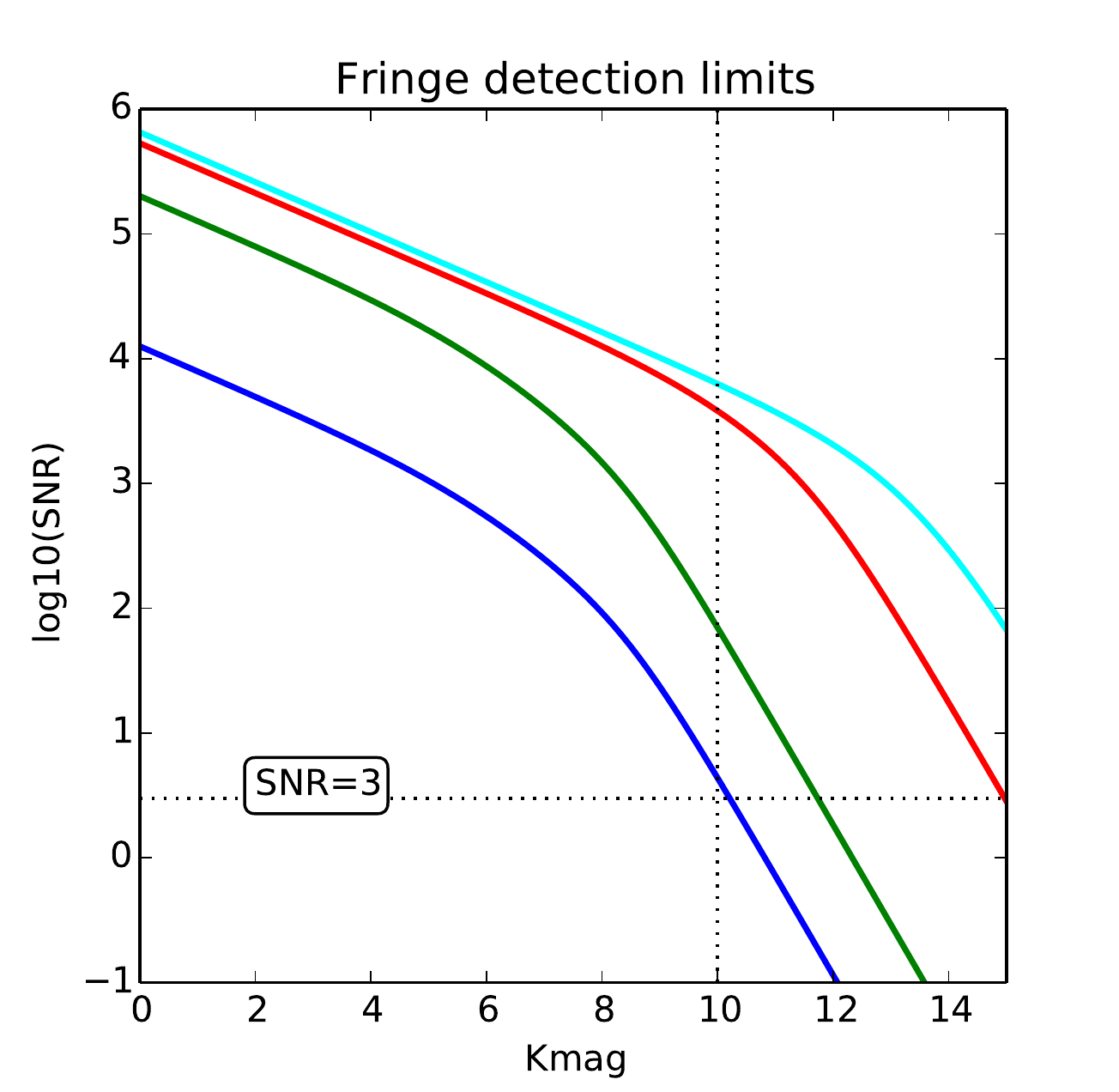}}
 \put(9, 0){\includegraphics[width=7cm, height=7cm]{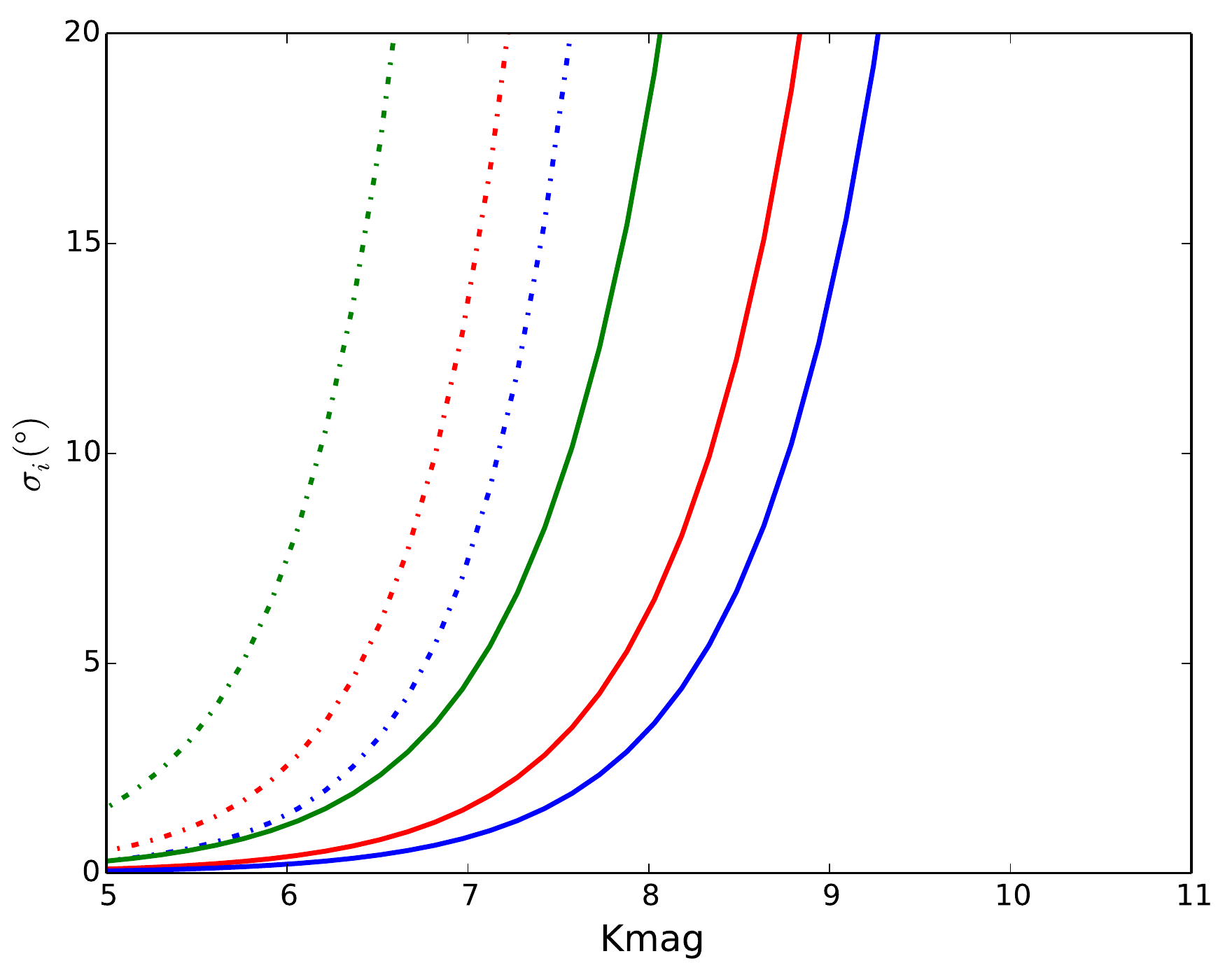}}
 
 \end{picture}
 
 \caption{Left: Fringe detection (log10(SNR)) limits for different VLTI instruments: from left to right: standard AMBER performance with 0.2 s frames (solid blue), current AMBER+ measured performance with incoherent TF2D processing (solid green), OASIS module (solid red) and OASIS+ module (solid cyan). The AMBER+ curve (given here for a maximum of 20s) is compatible with our experimental result of fringe detection with SNR=3 in 1 s on 3C273. The horizontal dotted black line shows the threshold fringe detection limit of SNR=3 and vertical dotted black line corresponds to $\mathrm{K}=10$. Right: Accuracy on inclination with K magnitude for AMBER+(dashed-dot) and GRAVITY(solid) for different inclination $i=5\degree$(green), $15\degree$(red) and $30\degree$(blue).}\label{Fig:SNR}
 
\end{figure}
   
 \subsection{VLTI INSTRUMENTS}
 We focus on the VLTI because we believe that the 8m UTs are a key feature for an AGN MR program. Table \ref{TABLE:1} summarizes the observing parameters for the different instruments. 
 
 {\bf AMBER:} AMBER \cite{2007A&A...464....1P} is the existing first generation near infrared spectro-interferometric VLTI instrument. With its standard frame-by-frame processing, it cannot observe AGNs in medium resolution (MR) without a fringe tracker that stabilize the fringes. The current VLTI fringe trackers are limited to about K$<9$. AMBER can be used in low resolution for absolute visibility measurements in the continuum.
 
{\bf AMBER+:} AMBER+ \cite{2012SPIE.8445E..0WP} refers to a new observing mode and data processing of AMBER data, as discussed in §\ref{sec:ref}. It allowed observing successfully the QSO 3C273 in MR (R=1500). The fringes were detected with a SNR=3 in typically 1s. To obtain differential visibility and phase with a sufficient accuracy (respectively 0.02 to 0.03 and $1\degree$ to $2\degree$) it has been necessary to bin the spectral channels down to a resolution 250. The results achieved with AMBER+ on 3C273 have been used to validate our SNR computations.

{\bf OASIS:} OASIS (“Optimizing AMBER for Spectro-Interferometry and Sensitivity”) is a low-cost module that could easily be installed on AMBER and use the AMBER+ software (Petrov, 2014). It has been designed to optimize the AMBER+ mode for medium spectral resolution observations. Its two main characteristics are: 1. It uses a spectral encoding to separate the fringe peaks, instead of the standard AMBER or MATISSE spatial encoding. Thus each interferogram can be coded on 4 pixels per spectral line instead of 32 like in AMBER (24 in GRAVITY). 2. The spatial filters with fibers are bypassed by optimized optics, allowing to use both polarizations together, which yields a gain in transmission of about 7 with regard to the current AMBER instrument.
   \begin{figure*}[!ht]
     \centering
      \setlength{\unitlength}{1cm}
       \begin{picture}(16, 11)
       \put(0, 0){\includegraphics[width=16cm, height=11cm]{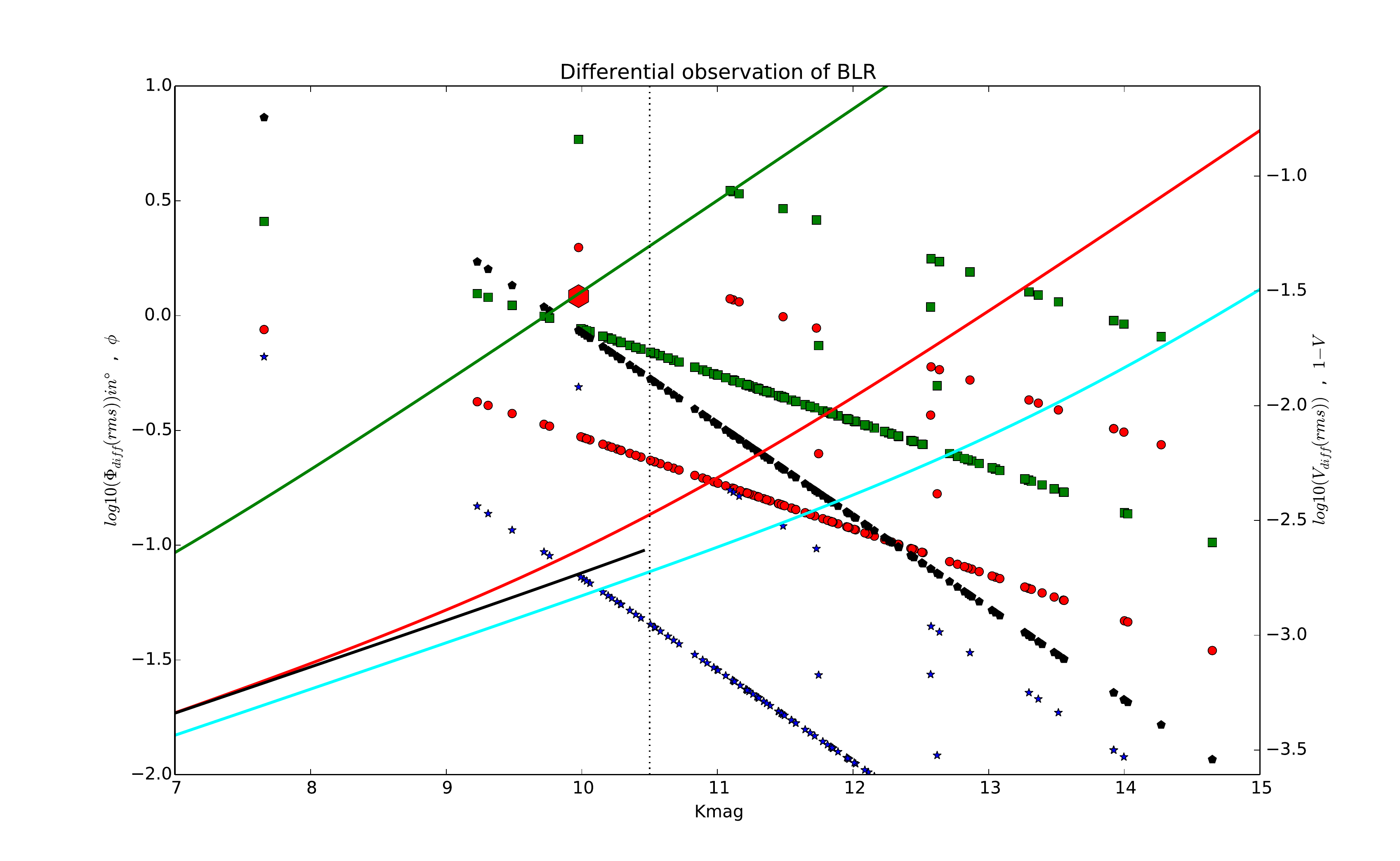}}
       \put(5.0, 8.5){3C273}
       \put(5.5, 8.4){\vector(1, -1){0.65}} 
       \put(7.5, 9.7){AMBER+}
       \put(12.0, 8.5){OASIS}
       \put(13.5, 7.5){OASIS+}
       \put(5.5, 2.5){GRAVITY}
       \put(6, 2.8){\vector(0, 1){0.7}}   
       \end{picture}
      \caption{Feasibility of observation of BLR with 135 m baseline: the lines represent the log of differential phase error  in deg (left side) and differential visibility error (right side). for the current AMBER+ processing (solid green), with the OASIS module (solid red), with the OASIS+ module (solid cyan) and with upcoming VLTI instrument GRAVITY (solid black). Each symbol represents one Sy 1 AGN observable at Paranal, from SIMBAD database; differential phase from skewed and inclined inner dust rings (filled green square), differential phase expected from $R_{blr}$ estimates from visible RM (filled red circle), differential visibility measurements (1-Vd) (filled blue star) and absolute visibility(1-$V_{abs}$) in black filled polygon. The vertical black dotted line shows the magnitude limit of GRAVITY.}\label{Fig:target}
   \end{figure*}
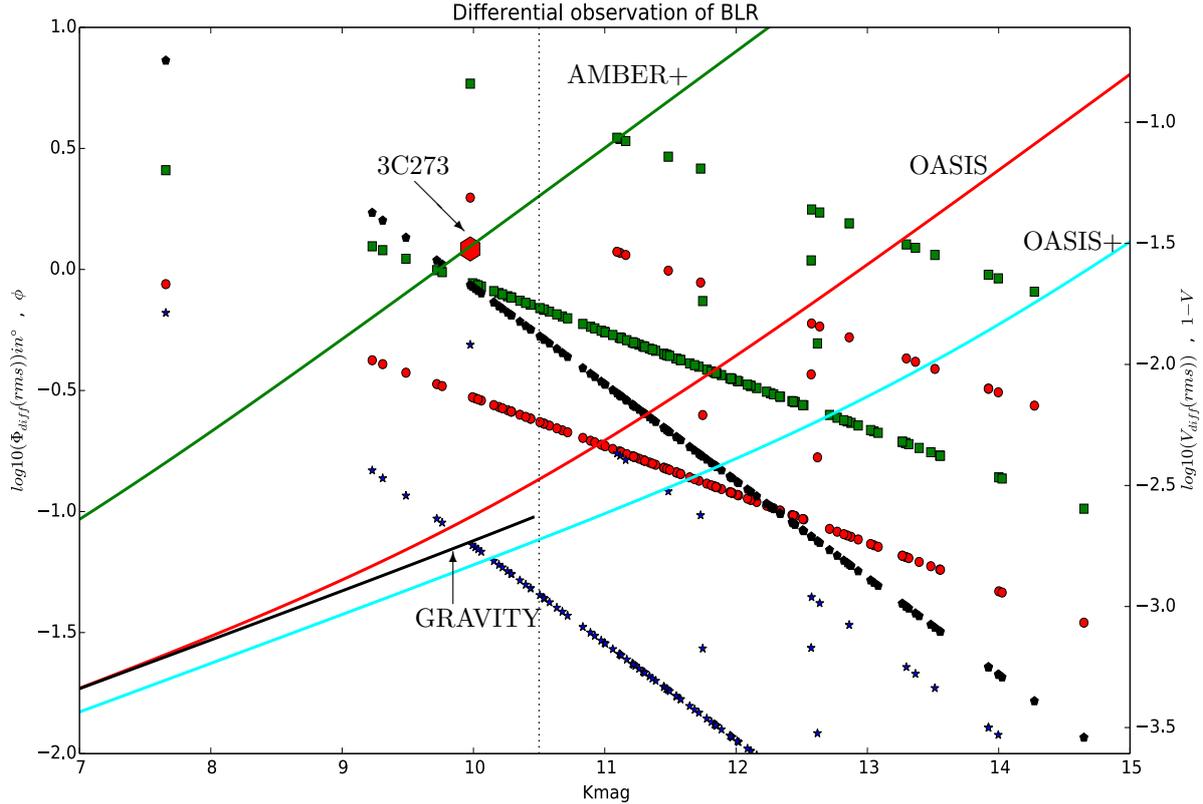
   
   {\bf GRAVITY:} GRAVITY \cite{2008SPIE.7013E..69E} is a 2nd generation VLTI spectro-interferometric instrument in K band. It should be commissioned in 2015-16. Its main characteristic of interest for a BLR program is that it has an internal fringe tracker that should allow cophased observations up to K=10.5. This allows much longer individual frame times, a higher instrumental visibility and a more stable one. The current GRAVITY plans do no foresee operation without its fringe tracker, but it should be possible to implement on it a GRAVITY+ TF2D mode allowing to observe beyond K=10.5, with performances intermediate between AMBER+ and OASIS.
 
 {\bf OASIS+:} OASIS uses the spectrograph and the detector of AMBER. OASIS+ would use OASIS with a new SELEX detector and a spectrograph optimized for BLRs, with a spectral resolution R=500. OASIS+, or any other successor of the 2nd generation VLTI instruments, is not in the current ESO plans, but it gives an idea of what could be ultimate VLTI performance for AGNs. 
  \begin{center}
   \begin{table*}[!hb]
  	\caption{Parameters for fringe detection limit$^a$ and differential observation of BLR$^b$ plot.}
  	\begin{minipage}{\textwidth}  
   	\centering {
   	\begin{small}
   	\tabcolsep=0.17cm
       \begin{tabular}{ |c |c| c| c| c| c| c| c|c| c| c| c| c| c|}  \hline
       \multicolumn{1}{ |c| }{ \bf{Instrument}} 
        & \multicolumn{13}{ c| }{ \bf{Parameters}} \\ \cline{2-14}
       \multicolumn{1}{|c|}{} & $n_T$ & $t_{DIT} (s)$ & $n_p$ & $\sigma_{RON}$ & $n_{th}$ & $V_{inst}$ & $n_\lambda$ & $N_{EXP}$ & $n_0$ & A($cm^2$) & S & T & R \\ \hline
       AMBER    & 3		& 0.2	& 64 & 	11	& 1.07	& 0.25 	&  256	& 100$^a$, 36000 $^b$ & 45	& 497628 & 0.5	& 0.02		& 1500   \\ \hline
       AMBER+   & 3		& 0.2	& 64 & 	11	& 1.07	& 0.25	&  256	& 100$^a$, 36000$^b$ & 45	& 497628 & 0.5	& 0.02		& 1500   \\ \hline
       OASIS    & 3		& 0.2	& 8	 &  11	& 1.07	& 0.25	&  256	& 100$^a$, 36000$^b$ & 45	& 497628 & 0.5	& 0.02 $\times$ 7	& 1500  \\ \hline
       OASIS+   & 3		& 0.1	& 8	 & 	3	& 1.07	& 0.5   &  256	& 100$^a$, 36000$^b$ & 45	& 497628 & 0.5	& 0.02 $\times$ 7	& 500   \\ \hline
       GRAVITY  & 4	    & 60  & 48\footnote[3]{Pair wise, 8 per pair} & 11 	& 1.07	& 0.75	&  256	& 120$^b$			 & 45	& 497628 & 0.5	& 0.01 $\times$ 2	& 500   \\ \hline      
   	\end{tabular} \label{TABLE:1}
   	\end{small}
   }
   \end{minipage}
   \end{table*}
   \end{center} 
 
 \subsection{Fringe detection limit}\label{sec:SNR}
 
 In left panel of Fig.\ref{Fig:SNR} we plotted the fringe detection limit log(SNR) with $\mathrm{K}$ mag. using the parameters listed in Table \ref{TABLE:1} for  different instrument like standard AMBER performance with 0.2 sec frames (blue), AMBER+ performance with incoherent TF2D processing (green), OASIS module (red) and OASIS+ module (cyan). We found that the AMBER+ could reach up to $\mathrm{K}\sim 10.5$ and the potential limit of the new OASIS and OASIS+ $>13$. The accuracy on the inclination with $\mathrm{K}$ mag. is plotted in right panel of Fig.\ref{Fig:SNR} for different instrument AMBER+ (dotted) and GRAVITY (solid) for various inclination $5\degree$(green), $15\degree$(red) and $30\degree$(blue).


 \subsection{Targets and amplitude of signals}\label{sec:target}
 
 We collected a list of all Sy1 and QSOs observable with the VLTI found in the SIMBAD catalog with specific search criteria $\mathrm{K}<13$, $\mathrm{V}<15$ and $\mathrm{dec}<15\degree$. For each source we estimate the inner rim radius from its magnitude thanks to an extrapolation of Ref.~\citenum{2006ApJ...639...46S} known measurements. From this rim radius we evaluate the possible values of the continuum visibility and the differential visibility and phase. These values are compared to the SNR estimates.
 
 We use the CMB corrected redshift for each target from NED. The $\mathrm{K}$ mag. of each object is corrected taking the $\mathrm{K}$ mag. from 2MASS point source catalog when available (i.e. for almost all source). We subtracted the contribution of host galaxy, which is taken 0.2 in $\mathrm{K}$ \cite{2004ApJ...600L..35M}. 
 
 We collected the list of objects from Ref.~\citenum{2013ApJ...767..149B} that has classical RM BLR size. Then we fitted the radius with their $\mathrm{K}$ mag. and extrapolate for the objects that do not have the RM BLR size. In a similar way we obtained dust sublimation radius from IR RM observation of Ref.~\citenum{2006ApJ...639...46S} and extrapolate for the objects that has no IR measurement.
 
 To estimate absolute visibility signal we used the following equation deduced from the appendix A. 
 \begin{equation}
 V_c=1-2\alpha_c^2,
 \end{equation}
 where $\alpha_c=\frac{2R_{rim}}{\lambda/B}$. 

Differential visibility yields the relative values of inner rim radius $R_{rim}$ \cite{2006ApJ...639...46S} and $R_{blr}$ \cite{2013ApJ...767..149B}. If $R_{rim}<<R_{blr}$, the differential visibility yields $R_{rim}$. If they are different and both smaller than $\lambda/B$, $R_{rim}$ and $R_{blr}$ can be deduced independently from measurements at different baselines. The differential visibility signal is typically
 
 \begin{align}
 V_{diff} & \simeq  - \frac{S_l}{S_l+S_c} \frac{\alpha_c^2}{1-\alpha_c^2} \quad \text{when  $R_{blr} << R_{rim} $}\\
          & \simeq  + 3 \frac{S_l}{S_l+S_c} \frac{\alpha_c^2}{1-\alpha_c^2} \quad \text{when  $R_{blr} =2R_{rim} $},
 \label{eq:v1}
 \end{align}
 where $S_l$ is the line strength and $S_c$ is the continuum strength for $R_{blr}$ ranging from 0 to $2R_{rim}$ and $S_c=1$.\\
 The typical differential phase amplitude for the BLR only is given by
 \begin{equation}
 \phi_{diff}=\pi \frac{S_l}{S_l+S_c}\alpha_{l}cos(\omega),
 \label{eq:phi1}
 \end{equation}
 where $\alpha_{l}=\frac{2R_{blr}}{\lambda/B}$ and $\omega$ is the opening angle of the disc.
 
 If the inner rim of the dust torus is inclined and skewed, differential interferometry will also be sensitive to the difference between the continuum apparent photocenter with maximum amplitude of
 \begin{equation}
 \phi_{diff}\simeq \frac{\pi}{2} \frac{S_l}{S_l+S_c} \alpha_c sini
 \label{eq:phi2}
 \end{equation}
 For line strength, we considered:
 \begin{itemize}
 \item $S_l=0.6$ when $\mathrm{Pa}\,\alpha$ is in the K band ($0.08\le z < 0.25$)
 \item $S_l=0.3$ when $\mathrm{Pa}\,\beta$ is in the K band ($0.4\le z<0.87$)
 \item $S_l=0.3$ when $\mathrm{Pa}\,\beta$ is in the H band ($0.25\le z<0.4$)
 \item $S_l=0.06$ when $\mathrm{Br}\,\gamma$ is in the K band ($z<0.08$)
 \item $S_l=0.12$ when $\mathrm{Pa}\,\gamma$ is in the K band ($z\ge 0.87$)
 \end{itemize}
 
The above eq. \ref{eq:v1}-\ref{eq:phi2} are certainly model dependent but they are sufficient for an estimation of the possibility to observe a target. We will use the above equation in §\ref{sec:present and future} to estimate the SNR from different interferometric instruments and the number of possible BLR of AGNs accessible by present and upcoming interferometers in VLTI. 

\subsection{Potential Performance}
 Fig.\ref{Fig:target} shows the feasibility of observation of BLRs with current and upcoming VLTI instruments like current AMBER+ processing (solid green), with the possible low cost OASIS module (solid red), the OASIS+ module (solid cyan) and next generation upcoming VLTI instrument GRAVITY (solid black). Each symbol represents one Sy 1 AGN observable at Paranal, from SIMBAD database; differential phase from skewed and inclined inner dust rings (filled green square) from eq.\ref{eq:phi2}, differential phase expected from $R_{blr}$ estimates from visible RM (filled red circle) from eq.\ref{eq:phi1} and differential visibility measurements (1-Vd) from eq.\ref{eq:v1} (filled blue star). Any signal with accuracy better than these amplitudes will yield useful constraints.
 
 We found that with the improved modules like OASIS and OASIS+ on AMBER we shall be able to access few objects where we can get all the differential interferometric measures, two times more objects shall provide differential phase signal even if the interferometric radius is as small as the BLR radius and asymmetry in the inclined skewed disk (or a BLR phase effect for an extended BLR) could be detectable with differential interferometry for three times more object. The upcoming instrument GRAVITY with its new fringe tracking capability up to $\mathrm{K}=9.5$ for Sy 1 AGNs and $\mathrm{K}=10.5$ for QSOs is likely to provide all interferometric measures on many of the targets.   
 
 {\it For a good estimation of continuum and BLR sizes we need accurate absolute visibility measurement. Differential visibility and phases with medium spectral resolution (R$>1000$) is necessary to constrain the geometry of the BLR. Differential phase alone, possibly assisted by lower spectral resolution differential visibility (in the 250-500 range) can provide size and global velocity estimates, i.e. masses if we have an a priori knowledge of the model.}



\section{Discussion}\label{sec:diss}

From our dynamical model, we showed that the differential visibility, due to its higher accuracy, allows accurate measurement of BLR size for a known model. If we also have the absolute visibility measurement, we can have an almost model independent size measurement. 
Our model and the SNR estimates show that the absolute visibility, differential visibility and differential phase at medium spectral resolution can strongly constrain the parameters of a fairly complex model. This could be implemented in a global model fitting way. The study of the corresponding parameter degeneracy and final accuracy would be an extension of the current work.


We found that the BH mass measurement is very sensitive to global velocity of the BLR clouds as well as the scale factor $f$ which depends on the geometry ($\omega$ and $i$) and local line width (here, $\sigma0_{line}$) that can represent the combination of microturbulence inside the clouds with macroturbulent motion of the clouds. The RM 1D response function produced in different combinations of these parameters are similar and hence RM can not discriminant these three parameters. However both the differential visibility and differential phase measures with a sufficient spectral resolution (typically $>1000$) allow to discriminate them. The effect is more obvious in the differential visibility but the differential phase can be measured accurately on smaller targets.  

An independent estimate of the inclination would very substantially improve the accuracy on $\omega$ and $\sigma0_{line}$. This estimates could be obtained from detail line profile fitting and radio emission from jets for individual objects though the uncertainty is too large and the necessary jet observations from the visible to the X-ray domain are not easily available on all targets. Based on the emission line fitting Ref.~\citenum{1994ApJS...90....1E} has found that the inclination of BLR is $24\degree-30\degree$ and Ref.~\citenum{1996ApJ...459...89E} suggested the inclination of the BLR can be $19\degree-42\degree$. X-ray studies of Sy 1 suggested that the BLR inclination should be low \cite{1997ApJ...477..602N, 1999ApJ...523L..17N}. Thus we should rely on the OI interferometric measures of the ratio of visibilities between the major and minor axis of the dust rim. This needs low spectral resolution observations of the K band continuum with accurate absolute visibility measurements. GRAVITY would give $\sim13 \degree$ accuracy on $30\degree$ inclined object for K=9, if we achieve an absolute visibility accuracy of 0.003 like the FLUOR instrument on CHARA. We expect that GRAVITY will achieve this kind of accuracy but this still has to be proven. Our work sets a specification for this measure for a full use of the AGN BLR program.


To allow more precise estimates of the mass-luminosity relation, which would be a major contribution to the study of SMBH and host galaxy evolution, we hope that it will be possible to calibrate a law linking the RM projection factor to the luminosity: $f_{RM}=f(L)$. This needs a calibration of the $\omega=f(L)$ and $\sigma0_{line}=f(L)$ laws.
The study of $\omega=f(L)$ needs differential visibilities and phases with a spectral resolution higher than 1000. If we access all targets permitted by VLTI instrument, as shown in Fig.\ref{Fig:target}, we cover a wide luminosity range $10^{43}$-$10^{48}$ ergs/s.



Combined measurement of BLR size of the same object with RM and OI will allow to measure the distance of QSO by “Quasar Parallax” according to Ref.~\citenum{2002ApJ...581L..67E}, however we need OI to constrain a model to be able to connect angular size from OI to the linear size estimated by RM. With the help of differential phase measurement we could constrain the interferometric size of the BLR upto redshift z$\simeq$1.7. 

The RM lag-luminosity relationship can also be very substantially improved by a better geometrical model. The potential gain will be studied in a next paper.

VLTI is the only interferometer that allows to observe AGN at medium spectral resolution thanks to its large apertures however its baseline is limited to 135m. The upcoming VLTI instrument GRAVITY will be commissioned in 2015-2016. Its impact on the AGN program critically depends on the performances of its internal fringe tracker, but the announced limiting magnitude of K=10.5 seems very reasonable and maybe even slightly conservative. OASIS module could be installed on AMBER in a few months, as soon as ESO accepts to include the corresponding extremely moderate workload in the VLTI planning. OASIS+ which is the AMBER improvement, dedicated to AGN observations, could be a visitor instrument in the 1M euros range and can be developed in less than two years, but, for management reasons, its installation on the VLTI must wait at least the full completion of the 2nd generation general user instruments GRAVITY and MATISSE, i.e. at least 2018. The possibilities of all these instruments, and in particular of GRAVITY and OASIS, can be boosted by the implementation of a new generation Fringe Tracker with a limiting magnitude larger than this of GRAVITY. Concepts allowing to reach the K=12 to K=13 range have been proposed \cite{2014SPIE.9146.P} and phase-A studies are at different completion degrees. If fast decisions are made, such a device could be available around 2018.

The ultimate goal would be full images with R=1000 though the emission lines. The angular resolution of BLR requires accurate closure phases and therefore kilometric baselines in the near infrared, i.e. the development of a new facility. It can also be achieved in the visible with the current CHARA baselines, but would require a medium resolution limiting magnitude in the V=13-14 with 1 m telescopes. In both cases, it is not yet possible to give a date for these achievements. 

\section{Conclusion}\label{sec:conclusion}
A large number of BLR, up to K$>13$, will be accessible with the upcoming VLTI instruments GRAVITY and possible future instrument OASIS, which will unveil the morphology of the BLR and hence it will make possible to answer the central SMBH growth and evolution history and would allow the test of a further unification step, based on the dependence of key parameters from the luminosity. New generation FT and OASIS+ (or another VLTI module optimized for AGNs), will allow using the full VLTI potential, as it is limited by its baselines. The future developments of this work include: A. The presentation and interpretation of the observation of 3C273 with AMBER+/VLTI. B. The analysis of a full model fitting approach, based on more detailed measurement accuracies achieved on recent 3C273 data. C. The physical modeling of clouds, using the cloudy model, and allowing to compute the clouds spectral response and their contribution to luminosity in interaction with the global geometrical and kinematic parameters. D. An analysis of the contribution of OI interferometry in the thermal infrared with the MATISSE VLTI instrument.

\appendix    

\section{Angular size and OI measures} \label{sec:app}
We can derive simple expressions of the visibility and the closure phase if we assume, without the loss of generality that the object has two components:
\begin{itemize}
\item[a)] A symmetric component, with total flux $1-a$ and equivalent radius $R$.
\item[b)] An asymmetric component, with total flux a ($a<1$) located at a position $P=pR$ (with $p<1$)
\end{itemize} 

For a simple evaluation, we shall derive the interferometric measures for the following example of such an object:
\begin{equation}
O(r, \lambda)=\frac{(1-a)\delta(r-R)}{2} + \frac{(1-a)\delta(r+R)}{2} + a\delta(r-P)
\end{equation}
Eq.\ref{eq:photo} can be expanded as  
\begin{align}
 \tilde{O}(\mathbf{u}, \lambda) & =  \nonumber
  1 - i 2\pi  \frac{\int\int{\mathbf{u.r} O(\mathbf{r}, \lambda)\,\mathrm{d^2}\mathbf{r}}}{\int\int{O(\mathbf{r}, \lambda)\, \mathrm{d^2}\mathbf{r}}}  
  - \frac{(2\pi)^2}{2}  \frac{\int\int{\mathbf{(u.r)^2} O(\mathbf{r}, \lambda)\,\mathrm{d^2}\mathbf{r}}}{\int\int{O(\mathbf{r}, \lambda)\, \mathrm{d^2}\mathbf{r}}} 
  +i \frac{(2\pi)^3}{6}  \frac{\int\int{\mathbf{(u.r)^3} O(\mathbf{r}, \lambda)\,\mathrm{d^2}\mathbf{r}}}{\int\int{O(\mathbf{r},\lambda)\, \mathrm{d^2}\mathbf{r}}} +...\\  
  {}& =  1- iUM_1 + \frac{UM_2}{2} + i\frac{UM_3}{6} + ... ,
\label{eq:expan}
\end{align}
where $U=2\pi\mathbf{u}$ and $M_n=\int\int(\mathbf{r}^n O(\mathbf{r}, \lambda)d^2r)$ is the nth order moment of the brightness distribution $O(\mathbf{r},\lambda)$. The moments of the example brightness distribution are:
\begin{equation}
M_n = (1-a)R^n + a(PR)^n 
    =  a(PR)^n  
    =  1   
\end{equation}

Using the above values in eq.\ref{eq:expan} and doing some further calculation after neglecting the higher order terms of $UR$ we obtain:
\begin{align}
\phi_{*ij}(\lambda) & =-U\epsilon(\lambda) + \frac{U^3}{6} R^2\epsilon^2(\lambda)\left[3(1-a)+3ap^2(\lambda)-p^2(\lambda) \right]\\
\Psi_{123} & =-\frac{\epsilon(\lambda) R^2}{6} \left[3(1-a)+3ap^2(\lambda)-p^2(\lambda) \right] \times
\left(U_1^3 + U_2^3 + U_3^3\right) \\
V_{*ij} & =1-\frac{U^2R^2}{2}\left[(1-a)+ap^2(\lambda) \right] +\frac{\epsilon^2(\lambda)U^2}{2},
\end{align}
where $\epsilon(\lambda)=aPR$.
 
For non-resolved sources, we note $\alpha=\frac{\Lambda}{\lambda/B}=\frac{2R}{\lambda/B}>UR$.

\acknowledgments     
 
This research has made use of the SIMBAD database which is operated at CDS, Strasbourg, France and the NASA/IPAC Extragalactic Database (NED) which is operated by the Jet Propulsion Laboratory, California Institute of Technology, under contract with NASA. This publication also makes use of data products from the Two Micron All Sky Survey, which is a joint project of the University of Massachusetts and the Infrared Processing and Analysis Center/California Institute of Technology, funded by the National Aeronautics and Space Administration and the National Science Foundation. SR thanks Neha Sharma for providing criticisms to improve this manuscript.  

SR is supported by the Erasmus Mundus Joint Doctorate Program by Grant Number  2011-1640 from the EACEA of the European Commission.

\bibliographystyle{spiejour}   
\bibliography{REF}   

\end{document}